\documentstyle[12pt,aaspp4]{article}

\newcommand{\be}{\begin{equation}}
\newcommand{\ba}{\begin{eqnarray}}
\newcommand{\ee}{\end{equation}}
\newcommand{\ea}{\end{eqnarray}}
\newcommand{\bm}[1]{\mbox{{\boldmath $#1$}}}

\begin{document}

\begin{flushright}
KUNS-1616
\end{flushright}

\title{\bf On the Decelerating Shock Instability of Plane-Parallel Slab
with Finite Thickness}

\author{RYOICHI NISHI$^1$ and HIDEYUKI KAMAYA $^1,^2$}

\affil
{1 : Department of Physics, Kyoto University, Sakyo-ku, 
Kyoto 606-8502, JAPAN}
\affil
{2 : Department of Astronomy, Kyoto University, Sakyo-ku,
Kyoto 606-8502, JAPAN}

\affil{email(RN):nishi@tap.scphys.kyoto-u.ac.jp}
\affil{email(HK):kamaya@kusastro.kyoto-u.ac.jp}

\centerline{ Apr.~16th, 1999}
 
\begin{abstract}
 
Dynamical stability of the shock compressed layer with finite thickness
is investigated.
It is characterized by self-gravity, structure, and
shock condition at the surfaces of the compressed layer.
At one side of the shocked layer, 
its surface condition is determined via the ram pressure,
while at the other side the thermal pressure supports its structure.
When the ram pressure dominates the thermal pressure,
we expect deceleration of the shocked layer.
Especially, in this paper, we examine 
how the stratification of the decelerating layer
has an effect on its dynamical stability.
Performing the linear perturbation analysis, 
a {\it more general} dispersion relation than the previous one
obtained by one of the authors is derived.
It gives us an interesting information about
the stability of the decelerating layer;
When the deceleration of the layer dominates the self-gravity,
the so-called DSI (Decelerating Shock Instability) is efficient, 
while the gravitational instability
occurs when the self-gravity works better than the deceleration.
The length-scales of the two instabilities are the order of 
the width of the  decelerating layer.
Their growth time-scales are 
the order of the free-fall time in the density of the shocked slab. 
Importantly, they are always incompatible. 
We also consider the evolution effect of the shocked layer.
In the early stages of its evolution,
only DSI occurs. On the contrary, in the late stages,
it is possible for the shocked layer
to be unstable for the DSI (in smaller scale) and
the gravitational instability (in larger scale), ordinary.
The onset-time of the gravitational instability is 
the order of the free-fall time of the external medium. 
After onset, as shown above,
growth time-scale becomes the free-fall time 
of the dense slab, which is much shorter than the free-fall time 
of the external medium.
Furthermore, we find there is a stable range of wavenumbers
against both the DSI and the gravitational instability
between respective unstable wavenumber ranges.
These stable modes suggest the ineffectiveness 
of DSI for the fragmentation of the decelerating slab.
Thus, in various cases,
the structure of the shocked layer should be determined
when its stability is discussed. 

\end{abstract}
 

\section{INTRODUCTION}

The hydrodynamic shock waves are expected to play important roles in
many astrophysical phenomena, especially in the formation of the
structure. The shock waves are induced by supernovae (e.g. Tomisaka
1990), expanding H$_{\rm II}$ regions (e.g. Elmegreen 1989), 
molecular outflows around protostars (e.g. Nakano, Hasegawa, \& Norman
1995), stellar winds from early type stars (e.g. Stevens, Blondin, \&
Pollock 1992), and so on.
These expanding shock waves sweep up the external
matter and the swept matter will become the materials of the structure
at the next stage (e.g. Kimura \& Tosa 1988). 
Previously, it was postulated that 
such mechanism performs in the galaxy formation era (e.g. Ikeuchi 1981).
In this paper, 
paying attention to the finite thickness of the shocked layer
(cf. Vishniac \& Ryu 1989),
we reexamine its stability by linear perturbation analysis. 
Especially, we present a general dispersion relation including the effect
of the self-gravity of a shocked slab with finite thickness.

The stability of decelerating shock waves have been investigated,
especially in the last two decades. Vishniac (1983) firstly studies the
isothermal shock waves expanding spherically in a uniform medium with a
thin shell picture. Then, he has found that
it is overstable against linear perturbations and the smaller wavelength
perturbation is more unstable,
although plane shock waves in a steady state were believed to be
stable for rippling perturbation (Erpenbeck 1962).
His instability is attendant on the decelerating shock waves
(e.g. Nishi 1992).
Then, we shall call it DSI (Decelerating Shock Instability).
Importantly, after his series of papers (Ryu \& Vishniac 1987, 1988, 1991
Vishniac \& Ryu 1989) is published, Grun et al. (1991) has reported 
the existence of instability of shock wave in laboratory.
After that, Mac Low \& Norman (1993) study the nonlinear evolution
of the blast waves, and confirm Grun et al.. 
The self-gravity effect on non-linear stage is
examined numerically by Yoshida \& Habe (1992).
Dgani,  van Buren \& Noriega-Crespo (1996) also
discuss the stability of bow-shocks
by means of the numerical simulations, and the transverse acceleration
instability semi-analytically.
The evolution of the supernova remnant is discussed in
Chevalier \& Blondin (1995). 
Vishniac (1994) examines the stability of a slab bounded on both sides
by shocks, which is linearly stable but nonlinearly unstable.
His analysis is confirmed by numerical simulations 
(Blondin \& Marks 1996).
For the extended review, see Vishniac (1995).

There is an important astrophysical phenomenon
related to the decelerating shock waves. 
Elmegreen and Lada (1977) propose a 
scenario on sequential formation of OB star subgroups. 
According to their scenario, an H$_{\rm II}$ region 
formed by a group of OB stars expands in a 
molecular cloud and a shock wave sweeps up the gas. 
Afterwards, the matter swept up by the shock becomes 
gravitationally unstable and fragmentation occurs. 
Elmegreen and Elmegreen (1978) 
studies the stability of 
the isothermal layer supported by the thermal pressure 
on both sides and 
showed that the layer is gravitationally unstable 
and the growth rate has the maximum for the perturbation wavelength 
several times the thickness of the layer. 
The case of the high external pressure to the self-gravity
is examined by Lubow \& Pringle (1993). 
Welter (1982) calculates the growth rate of the perturbation in 
the layer which is supported 
by thermal pressure on one side 
and by balancing ram pressure of the shock on the other side. 
Voit (1988) 
investigates the incompressible sheet supported 
by unbalanced two thermal pressures and shows the coupling of 
the gravitational instability and the Rayleigh-Taylor instability. 
The effect of the magnetic field to the stability of 
the self-gravitating slab is precisely examined by 
Nagai, Inutsuka, \& Miyama (1998).

However, importantly, the shock-compressed layers 
around H$_{\rm II}$ regions are 
supported by the thermal pressure on one side and 
the ram pressure on 
the other side, and the two pressures are not always balanced. 
In this work, we expect that the instability related to a decelerating 
shock wave plays an important role. 
Moreover, we must study the instability of 
the shocked self-gravitating layers, 
since in the scenario of sequential star formation 
the next generation of OB stars is finally 
formed by self-gravity. 
Elmegreen (1989) calculates the evolution of an isothermal 
shocked layer and 
the growth of the perturbations in the evolving dense layer 
with self-gravity.
The estimated growth time-scale is $\sim 0.25 (G \rho_e)^{-1/2}$,
and is insensitive to the Mach number. Here $\rho_e$ is the
density of the external (preshock) medium.
On the contrary, Vishniac (1983) estimates 
a relatively short scale and small mass scale for
gravitational collapse, which are essentially determined by postshock 
density. Nishi (1992)'s estimate is resemble to Vishniac (1983). 
The discrepancy originates from the fact that they adopt different
approximation in their analyses each other. 
As pointed out by Mac Low \& Norman (1993),
one of the differences is that 
Vishniac (1983) assumes trans-sonic turbulence in the shell, while
Elmegreen (1989) assumes supersonic turbulence with Mach number close to
the shock velocity,
and Mac Low \& Norman's numerical work, in which gravity is neglected, 
may support Vishniac's assumption. 
Moreover, very importantly, the structure of the shocked slab are not 
incorporated in these analysis including gravity. 
Then, as presented just below,
we adopt another assumption of the shocked slab
to clarify its dynamical stability.

In the current paper, we investigate 
the instability of a plane-parallel slab with finite thickness, 
which slab is bound by thermal pressure on one side 
and by ram pressure of the shock on the other side. 
The pressures on the two sides are dissimilar. 
We assume incompressibility for the equation of state of 
the post shock layer (i.e. in the slab), 
though we assume compressible fluid 
at the shock surface. 
This assumption of incompressibility can be suitable, 
since highly compressed isothermal fluid behaves just like 
an incompressible fluid as shown by Elmegreen and Elmegreen (1978). 
Moreover, as demonstrated by Mac Low and Norman (1993),
this approximation may be indicated by the fact that
the DSI saturates with transonic transverse flows in the shocked slab. 
By this assumption we can obtain, in addition, an analytical solution 
to the perturbation equations, with which we can reveal the precise
physics of the instabilities.

In the following sections, 
we perform the linear perturbation analysis,
to study the stability of incompressible slab 
without self-gravity of the layer in $\S$ 2 and with 
self-gravity in $\S$ 3. 
We discuss the DSI and gravitational instability of incompressible slab 
in $\S$ 4. We also discuss the fragmentation process of the slab in $\S$ 5  
and summarize our results in the final section.

\section{The DSI of incompressible slab}

\def\vv{\bm{v}}
\def\ex{{\rm e}}

In order to investigate the stability of the slab of gas 
swept up by a shock front, we analyze an infinite, 
plane-parallel shock wave advances to the negative direction of $z$-axis 
with the velocity of $V_S$ in a uniform medium of density of $\rho_e$. 
We assume incompressibility of the gas except at the shock front 
and the density in the dense slab being $\rho_s$, 
and that the dense layer is confined from the rear by  
hot and tenuous gas with a finite pressure $P_t$. 
As first step, we ignore the self-gravity of the slab 
to present the dispersion relation of the pure DSI 
(cf. \S2 of Nishi 1992). 
We assume a coordinate that the unperturbed shock front 
is at $z = 0$. 
At the boundary $z = 0$, we impose shock boundary conditions: 

\be
[ \rho v _{\perp}] = 0, 
\ee
\be
[ P + \rho v _{\perp}^2 ] = 0,
~~~[ \rho v_{\perp} \vv_{\parallel} ] = 0
\ee
where $\rho$ and $P$ are the density and the pressure, and 
$v_{\perp}$ and $\vv_{\parallel}$ are the velocities of the fluid 
perpendicular and parallel to the shock surface, respectively.
The square brackets denote the difference of the enclosed 
quantity across the shock front. 

The fluid equations for a general first-order perturbation in the slab
are 

\be
{d \vv_1 \over d t} 
= - \nabla \left( {P_1 \over \rho_s} \right),         \label{four}
\ee
\be
- \nabla \cdot ( \rho_s \vv_1 ) = 0,                  \label{five}
\ee
where $P$ is the pressure of the fluid, 
and the subscripts 0 and 1 refer to unperturbed and 
perturbed quantities, respectively. 
All perturbed quantities are proportional to 
$e^{i ( k x + \omega t )}$, then
we can rewrite equations (\ref{four}) and (\ref{five}) as 

\be
\left\{ i \omega + ( u \partial_z ) \right\} v_{1 x} 
- i k \left( {P_1 \over \rho_s} \right) = 0, 
\ee
\be
\{ i \omega + ( u \partial_z ) \} v_{1 z} 
- \partial_z \left( {P_1 \over \rho_s} \right) = 0, 
\ee
\be
i k v_{1 x} + \partial_z v_{1 z} = 0, 
\ee
where $u ( = ( \rho_e / \rho_s ) V_S ) $ is $v_0$ in the slab 
and $v_{1 x}$ and $v_{1 z}$ are the $x$- and $z$- components of 
$\vv_1$, respectively. 

According to Vishniac \& Ryu (1989), we get the following 
three (approximate) solutions with two different assumptions.
For first set of them, we assume that $u \partial_z \ll i \omega$. 
Then, we can ignore the terms in parentheses and 
find a combination of the two independent solutions given by 

\be
P_1 = P_+ \ex^{k z} + P_- \ex^{- k z}, 
\ee
\be
v_{1 x} 
= - {k \over \omega} {P_+ \over \rho_s} \ex^{k z} 
  - {k \over \omega} {P_- \over \rho_s} \ex^{- k z}, 
\ee
\be
v_{1 z} 
= {i k \over \omega} {P_+ \over \rho_s} \ex^{k z} 
  - {i k \over \omega} {P_- \over \rho_s} \ex^{- k z}. 
\ee

To find third solution, 
we consider the case that
the perturbed quantities have large gradient and
$u \partial_z$ terms are important.
By inspection the third solution is 

\be
{P_1 \over \rho_s} = 0, 
\ee
\be
v_{1 x} = A \exp \left[ - {i \omega \over u} z \right], 
\ee
\be
v_{1 z} = {A u k \over \omega} 
      \exp \left[ - {i \omega \over u} z \right]. 
\ee
Since we are interested in only the case that perturbation grows, 
we assume that $- i \omega_I \gg H / u$ where
$H$ is a thickness of the shocked slab and $- i \omega_I$ means
a growth rate of the perturbation. 

We shall determine the boundary conditions. 
Assuming the very high compressibility at the shock surface,
we can suppose $P_S = \rho_e V_S^2$.
Thus, for the pressure boundary conditions, we integrate the equation 
of hydrostatic equilibrium (Goldreich \& Lynden-Bell 1965), 
and find them to the first order; 

\be
P_1 ( 0 ) = ( 1 - \beta ) {P_S \over \sigma_0} \rho_s \eta 
            - 2 \rho_e V_S v_{1z} ( 0 ), 
\ee
\be
P_1 ( H ) = ( 1 - \beta ) {P_S \over \sigma_0} \rho_s \zeta, 
\ee
where
$\beta ~(\equiv P_t/P_S)$ is the pressure ratio at the surface of the slab, 
$\sigma_0$ is a surface density of the slab,
$\eta (\equiv z_{s 1} - 0)$ and 
$\zeta (\equiv z_{t 1} - H)$ are 
the displacements of the shock and 
trailing surfaces, respectively. 
With the shock boundary conditions, we set the following  boundary condition 
\be
v_{1x} ( 0 ) = {k \over \omega} V_S v_{1z} ( 0 ). 
\ee

Applying the boundary conditions, we have 
\be
P_+ + P_- \ex^{- 2 k H} 
= ( 1 - \beta ) {c^2 \over H^2 \omega^2} k H 
  \left( P_+ - P_- \ex^{- 2 k H} \right), 
\ee
\be
{P_+ \over \rho_s} + {P_- \over \rho_s} 
= ( 1 - \beta ) {c^2 \over H^2 \omega^2} k H 
  \left( {P_+ \over \rho_s} - {P_- \over \rho_s} - i A u \right), 
\ee
\be
\left( 1 - {c^2 k^2 \over \omega^2} \right) A 
= {i V_S k^2 \over \omega^2} 
  \left( {P_+ \over \rho_s} - {P_- \over \rho_s} \right), 
\ee
where $c^2 \equiv u V_S = \rho_{e} V_S^2 /\rho_{s}$. 

Then we find the dispersion relation as 
\be
\omega^2 = {c^2 \over H^2} 
\bigg[ {k^2 H^2 \over 2} \pm k H 
\bigg\{ {k^2 H^2 \over 4} - ( 1 - \beta ) \coth [ k H ] k H 
    + ( 1 - \beta )^2 \bigg\}^{ 1 / 2} \bigg].  \label{threeeight}
\ee
The numerical estimate of (\ref{threeeight}) is presented in
figure 1, where the growth rate is normalized by $c/H$, 
that is, $\Omega = \omega / (c/H)$.
The wave-number is expressed as $K = kH$ and
$\beta = 0.5$ is adopted in the same figure.
As clearly shown in this figure, 
we can find that there is a set of DSI-mode (the other set is damping
mode).

We evaluate the limiting case of long wave length 
($k H \rightarrow 0$). 
In the case of $\beta (1 - \beta) \not= 0$, 
we have the growth rate as 
\be
{\Gamma \over c / H} \simeq 
2^{- 1 / 2} \beta^{1 / 4} (1 - \beta)^{1 / 4} (k H)^{1 / 2}. 
\ee
This growth rate is equal to the long wave length limit of 
equation (2.35) of Nishi(1992) with the assumption of 
$H = \sigma_0 / \rho_s$.

\section{The DSI of incompressible 
slab with self-gravity}

With the effect of self-gravity,
the equation of fluid motion for a general first-order perturbation is 

\be
{d \vv_1 \over d t} 
= \nabla \left( \psi_1 - {P_1 \over \rho_s} \right),   \label{gfour} 
\ee
where $\psi_1$ is a perturbed component of the gravitational potential.
All perturbed quantities are also set to be  proportional to 
$e^{i ( k x + \omega t )}$ in this section.   
Then we can rewrite equations with $\psi_1$ as 
\be
\left\{ i \omega + ( u \partial_z ) \right\} v_{1 x} 
+ i k \left\{ \psi_1 - {P_1 \over \rho_s} \right\} = 0, 
\ee
\be
\{ i \omega + ( u \partial_z ) \} v_{1 z} 
+ \partial_z \left\{ \psi_1 - {P_1 \over \rho_s} \right\} = 0, 
\ee
\be
i k v_{1 x} + \partial_z v_{1 z} = 0. 
\ee
As performed in the previous section,  
we assume that $u \partial_z \ll i \omega$ firstly. 
Namely, we ignore the advection terms 
then find a combination of the two independent solutions given by 

\be
\psi_1 = \psi_+ \ex^{k z} + \psi_- \ex^{- k z},       \label{gten}
\ee
\be
P_1 = P_+ \ex^{k z} + P_- \ex^{- k z}, 
\ee
\be
v_{1 x} 
= - {k \over \omega} 
     \left( {P_+ \over \rho_s} - \psi_+ \right) \ex^{k z} 
  - {k \over \omega} 
     \left( {P_- \over \rho_s} - \psi_- \right) \ex^{- k z}, 
\ee
\be
v_{1 z} 
= {i k \over \omega} 
     \left( {P_+ \over \rho_s} - \psi_+ \right) \ex^{k z} 
  - {i k \over \omega} 
     \left( {P_- \over \rho_s} - \psi_- \right) \ex^{- k z}. 
\ee

Next, since we are concerned for
the large gradient of perturbed quantities,
we must treat the $u \partial_z$-terms.  
By inspection, this third solution is found as

\be
{P_1 \over \rho_s} - \psi_1 = 0, 
\ee
\be
v_{1 x} = A \exp \left[ - {i \omega \over u} z \right], 
\ee
\be
v_{1 z} = {A u k \over \omega} 
      \exp \left[ - {i \omega \over u} z \right]. 
\ee
Since we are interested in only the case that perturbation grows, 
we assume that $- i \omega_I \gg H / u$ again. 

We must determine the boundary conditions. 
We have the perturbed column densities induced 
by the displacement of the surface as 

\be
\sigma_s = - \rho_s \eta \ex^{- k z} \ex^{i ( k x + \omega t )}. 
\ee
\be
\sigma_t = \rho_s \zeta \ex^{- k z} \ex^{i ( k x + \omega t )}, 
\ee
Since the fluid is incompressible, 
we can determine potential perturbation, $\psi_1$, 
from the surface density distributions 
$\sigma_s$ and $\sigma_t$ alone. 
With the constraints $\eta$, $\zeta \ll 2\pi /k$, 
we have the potentials due to $\sigma_s$ and $\sigma_t$ as 

\be
\psi_{\sigma_s} = - {2 \pi G \rho_s \over k} \eta 
                \ex^{- k z} \ex^{i (k x +\omega t)}, \label{gonenine}
\ee
\be
\psi_{\sigma_t} = {2 \pi G \rho_s \over k} \zeta 
           \ex^{k ( z - H )} \ex^{i (k x +\omega t)}. \label{goneten}
\ee
{}From the equations (\ref{gten}), (\ref{gonenine}), 
and (\ref{goneten}) we find 

\be
\psi_+ = {2 \pi G \rho_s \over k} \zeta \ex^{- k H}, 
\ee
\be
\psi_- = {2 \pi G \rho_s \over k} \eta. 
\ee

For the pressure boundary conditions, we find, to first order, 

\be
P_1 ( 0 ) = - 2 \pi G \rho_s^2 H \eta 
            + ( 1 - \beta ) {P_S \over \sigma_0} \rho_s \eta 
            - 2 \rho_e V_S v_{1z} ( 0 ), 
\ee
\be
P_1 ( H ) = 2 \pi G \rho_s^2 H \zeta 
            + ( 1 - \beta ) {P_S \over \sigma_0} \rho_s \zeta. 
\ee

Applying the shock boundary conditions gives the next boundary condition 
\be
v_{1x} ( 0 ) = {k \over \omega} V_S v_{1z} ( 0 ). 
\ee

Applying the boundary conditions, we get 
\be
\psi_+ = {2 \pi G \rho_s \over \omega^2} 
  \left\{ \left( {P_+ \over \rho_s} - \psi_+ \right) 
        - \left( {P_- \over \rho_s} - \psi_- \right) \ex^{- 2 k H} 
                                            \right\}, \label{gtwosix}
\ee
\be
\psi_- = - {2 \pi G \rho_s \over \omega^2} 
  \left\{ \left( {P_+ \over \rho_s} - \psi_+ \right) 
        - \left( {P_- \over \rho_s} - \psi_- \right) - i A u 
                                                    \right\}, 
\ee
\be
{P_+ \over \rho_s} + {P_- \over \rho_s} \ex^{- 2 k H} 
= ( \tilde \alpha + 1 ) {2 \pi G \rho_s k H \over \omega^2} 
  \left\{ \left( {P_+ \over \rho_s} - \psi_+ \right) 
        - \left( {P_- \over \rho_s} - \psi_- \right) \ex^{- 2 k H} 
                                                    \right\}, 
\ee
\be
{P_+ \over \rho_s} + {P_- \over \rho_s} 
= ( \tilde \alpha - 1 ) {2 \pi G \rho_s k H \over \omega^2} 
  \left\{ \left( {P_+ \over \rho_s} - \psi_+ \right) 
        - \left( {P_- \over \rho_s} - \psi_- \right) - i A u 
                                           \right\},  \label{gtwonine}
\ee
\be
\left( 1 - {c^2 k^2 \over \omega^2} \right) A 
= {i V_S k^2 \over \omega^2} 
  \left\{ \left( {P_+ \over \rho_s} - \psi_+ \right) 
- \left( {P_- \over \rho_s} - \psi_- \right) \right\}, \label{gtwoten}
\ee
where 
$\tilde \alpha ( \equiv ( 1 - \beta ) P_S / ( 2 \pi G \sigma_0^2 ) )$ 
is the ratio of the deceleration of the whole slab to 
the acceleration of the slab's self-gravity at the surface. 

{}From equations (\ref{gtwosix}) to (\ref{gtwonine}), we have 
\ba
B_+ + B_- &=& ( \tilde \alpha - 1 ) {2 \pi G \rho_s k H \over \omega^2} 
            ( B_+ - B_- - i A u ) \cr
          &-& {2 \pi G \rho_s \over \omega^2} 
            ( B_- - B_- \ex^{- 2 k H}  + i A u),     \label{gthreeone}
\ea
\ba
B_+ + B_- \ex^{- 2 k H} &=& ( \tilde \alpha + 1 ) 
            {2 \pi G \rho_s k H \over \omega^2} 
            ( B_+ - B_- \ex^{- 2 k H} ) \cr
          &-& {2 \pi G \rho_s \over \omega^2} 
            ( B_+ - B_+ \ex^{- 2 k H}  + i A u),     \label{gthreetwo}
\ea
where $B_+ \equiv P_+ / \rho_s - \psi_+$ and 
$B_- \equiv P_- / \rho_s - \psi_-$, respectively. 
Then we can rewrite equations (\ref{gthreeone}) 
and (\ref{gthreetwo}) as 
\ba
&{}&\bigg\{ {\omega^2 \over 2 \pi G \rho_s} 
      - ( \tilde \alpha - 1 ) k H \bigg\} B_+ \cr
&+& \bigg\{ {\omega^2 \over 2 \pi G \rho_s} 
      + ( \tilde \alpha - 1 ) k H 
      + ( 1 - \ex^{- 2 k H} ) \bigg\} B_- \cr
&+& \bigg\{ 1 + (\tilde \alpha - 1 ) k H \bigg\} i A u = 0, 
                                                   \label{gthreethree}
\ea
\ba
&{}&\bigg\{ {\omega^2 \over 2 \pi G \rho_s} 
        - ( \tilde \alpha + 1 ) k H 
        + ( 1 - \ex^{- 2 k H} ) \bigg\} B_+ \cr
&+& \ex^{- 2 k H} \bigg\{ {\omega^2 \over 2 \pi G \rho_s} 
        + ( \tilde \alpha + 1 ) k H \bigg\} B_- 
+ \ex^{- 2 k H} i A u = 0, 
\ea
and equation (\ref{gtwoten}) as 
\be
c^2 k^2 B_+ - c^2 k^2 B_- + ( \omega^2 - c^2 k^2 ) i A u = 0. 
                                                 \label{gthreefive} 
\ee

Thus, from equations (\ref{gthreethree}) $\sim$ (\ref{gthreefive}), 
we have dispersion relation
\ba
\omega^6 &+& 2 \pi G \rho_s 
   \left\{ 2 - {\tilde \alpha \over 1 - \beta} k^2 H^2 
             - 2 \coth [ k H ] k H \right\} \omega^4 \cr
&+& ( 2 \pi G \rho_s )^2 \bigg\{( \tilde \alpha + 1 ) 
         { \tilde \alpha \over 1 - \beta} \coth [ k H ] k^3 H^3 \cr
         &-& {\tilde \alpha \over 1 - \beta} k^2 H^2 
          - ( \tilde \alpha^2 - 1 ) k^2 H^2 \cr
&{}&- 2 k H + \left( 1 - \ex^{- 2 k H} \right) \bigg\} \omega^2 
= 0. 
\ea

We neglect the mode of $\omega = 0$, 
because of the assumption of $- i \omega_I \gg H / u$. 
Finally, then,  we obtain the dispersion relation : 
\ba
{\omega^2 \over 2 \pi G \rho_s}
&=& {\tilde \alpha \over 1 - \beta} {k^2 H^2 \over 2} 
  + \coth \left[ k H \right] k H - 1 \cr
&\pm & \bigg\{ \left( {\tilde \alpha \over 1 - \beta} {k^2 H^2 \over 2} 
  + \coth \left[ k H \right] k H - 1 \right)^2 \cr
&{}&  - ( 1 + \tilde \alpha ) {\tilde \alpha \over 1 - \beta} 
    \coth \left[ k H \right] k^3 H^3 \cr
&{}&+ {\tilde \alpha \over 1 - \beta} k^2 H^2
  - ( 1 - \tilde \alpha^2 ) k^2 H^2 + 2 k H 
  + \left( {\rm e}^{- 2 k H} - 1 \right) \bigg\}^{1 / 2}. 
                                               \label{gthreeseven}
\ea

\section{Discussions}

First,  
we investigate the limiting cases of the dispersion relation 
of equation (\ref{gthreeseven}). 
To the limit of the long wavelength, that is $k H \rightarrow 0$, 
we have 
\be
{\omega^2 \over 2 \pi G \rho_s} \simeq 
\pm k H ( 1 - \alpha^2)^{1 / 2}, 
~\alpha \equiv 
\left( {\beta \over 1 - \beta} \right)^{1 / 2} \tilde \alpha .
                                                   \label{gthreeeight}
\ee
{}From equation (\ref{gthreeeight}), 
we can classify the characteristics of the instability by $\alpha$.
In the case of $\alpha > 1$, the layer is overstable for 
long wavelength perturbation. 
When the self-gravity of the slab
does not play important roles, that is,
in the limits of $\tilde \alpha$ and $\alpha \rightarrow \infty$,
we have the dispersion relation of (\ref{threeeight}).
On the other hand, in the case of $\alpha < 1$, the layer is 
unstable for gravity modes. 

The numerical presentation of (\ref{gthreeseven}) is given in
figures 2-4, where the growth rate is normalized 
by $(2\pi G \rho _s)^{0.5}$,
that is, $\Omega = \omega / (2\pi G \rho _s)^{0.5} $.
The wave-number is expressed as $K = kH$ in all the three figures.
In figure 2, we display the cases of $\alpha = 1.2$  
and $\beta =0.2$, 0.4, 0.6 and 0.8. 
Similarly to figure 1, thus, we find the DSI dispersion relations
in figure 2. 
For the same value of $\alpha $,
$\beta$ is the larger, the unstable region becomes
the narrower and the growth rate becomes the smaller.
This is because large $\beta $ means inefficiency of the deceleration at
the surface of the slab.
In the case of the adopted parameters,
there is not  instability due to the self-gravity.

In figure 3, the dispersion relations of $\alpha = 0.7$
and $\beta $=0.2, 0.4, 0.6 and 0.8 are presented.
Since $\alpha $ is smaller than unity,
we expect the gravitational instability in a range of small wavenumbers.
Indeed, we find the gravitational instability for all $\beta $
in a range of small $K$.
As indicated in this figure,
the growth rate of the gravitational instability
depends almost only on $\alpha$.
On the other hand, we also obtain the DSI in larger $K$
except the case of $\beta=0.8$.
We find again that the growth rate of
the DSI does not depend only on $\alpha$ but also on $\beta$. 
By the way,
we want to comment here that for all of the three cases
($\beta =0.2$, 0.4 and 0.6) is stable
in the middle of $K$ in figure 3,
that is, both of the DSI and gravitational instability
do not occur for a special range of wave-numbers.
For the case of $\beta=0.8$, moreover, the DSI mode is always
stable because of small $\tilde \alpha$.
These points are the result
of the analysis incorporating the structure of the shocked slab, then,
it does not appear in Nishi (1992).

In figure 4, we present the dispersion relation for $\tilde \alpha 
= \alpha =0$.
As clearly shown in the solid lines, there is only gravitational
instability.
In other words, we get no DSI mode.
Thus, for the case of $\tilde \alpha, ~\alpha \rightarrow 0$,
the self-gravity controls the instability mainly.
Indeed, if  $\tilde \alpha, ~\alpha \rightarrow 0$, we obtain 
\be
\omega^2 = 2 \pi G \rho_s 
\Bigg[ - ( 1 \pm {\rm e}^{- k H} ) 
       + {\left( {\rm e}^{k H / 2} \pm {\rm e}^{- k H / 2} \right)^2 
          \over {\rm e}^{k H} - {\rm e}^{- k H}} k H \Bigg]. 
\ee
This is the same dispersion relation 
given by Goldreich and Lynden-Bell (1965). 
Thus, so long as $\alpha$ is different from unity somewhat, 
the order of the growth rate of the gravitational instability becomes  
$(G \rho_s)^{-1/2}$ and the scale-length of the most unstable wavelength 
becomes $H$ (see also figure 3), which are almost in agreement with 
the result of Elmegreen and Elmegreen (1978)'s high pressure case. 
To obtain this mode, 
it is important to incorporating the structure of the slab. 

Next, we display the linear displacements, $\eta $ and $\zeta$,
each of which means $\eta = z_{s 1} - 0$ and
$\zeta = z_{t 1} - H$, respectively. 
Figure 5 shows evolution of the linear displacements for the case of
$\alpha \gg 1.0$ and $K = 0.24$.
The solid line is $ z_{t 1}$ and the dotted line is $z_{s 1}$.
{}From the upper to lower panels,
we display the evolution of the shocked slab 
at (a)$0.0$, (b)$1.0 \cdot \delta t$, and (c)$2.0 \cdot \delta t$  
where $\delta t$ is $\pi / 4 |\omega (k = 0.24H)|$.
The unit of the length is $H$.
We find the bending property
of the  shocked slab for DSI clearly. 
On the other hands, figure 6 displays the evolution of the gravitational
instability of the shocked slab ($\tilde \alpha, ~\alpha \to 0$ ).
Each of the displayed time is 
(a)$0.0$, (b)$1.0 \cdot \delta t$, and (c)$2.0 \cdot \delta t$,
respectively.
The one set of evolution is very different from that of DSI.
Someone might think that the shocked slab
has evolved nonlinearly. Indeed, the final panels of Fig.5 and Fig.6
shows difficulty of the description of the linear stage.
We present both figures only for reader's clear inspection.
The essential results of the current discussions are derived from
figures 1-4.

Previously, Yoshida \& Habe (1992) discussed the nonlinear stage of DSI
and gravitational instability under the similar condition adopted 
in this paper.
According to them, the gravitational instability has longer wave-length
than DSI. This property is understood through our linear analysis,
since the both modes are incompatible in the dispersion relation
derived in (\ref{gthreeseven}).
Thus, our result is in accord with the calculation given by Yoshida \& Habe.
Someone might think that DSI does not have bending shocked layer
since we can not find it clearly in figure 9 of Yoshida \& Habe (1992).
However, although the thickness of our slab is nearly constant
in our linear analysis, 
in the numerical results of Yoshida \& Habe, 
the shocked slab has a few times larger thickness than the initial value.
If the bounding effect were to be more tight in Yoshida \& Habe (1992),
they would obtain our bending property of DSI in their simulations
(cf. Chevalier \& Theys 1975).

\section {Fragmentation of the shocked slab}

In the previous considerations, 
we assume that $\alpha$ is always constant in each case.
But, in reality, $\alpha$ changes with the evolution of the shocked slab 
(e.g., Elmegreen 1989).  In this section, then,
let's discuss the evolution effects (e.g., Nishi 1992).
{}From the definition,  we find
$\alpha \propto P_S / \sigma_0^2 \propto (V_S / \sigma_0)^2$.
Since the shocked layer sweeps up the matter,
the column density increases and shock velocity decreases with time,
hence $\alpha$ decreases.
Thus, the character of the instability in the slab changes at the time
$t=t_{\rm cr}$ when $\alpha$ becomes unity.
In the early stages
of the evolution of the shocked layer, $t < t_{\rm cr}$,
only the DSI can occur in that layer.
On the contrary, in the late stages, $t > t_{\rm cr}$,
it is possible for the shocked layer
to be unstable for the DSI (in smaller scale) and
the gravitational instability (in larger scale), ordinary.
Finally, $t \gg t_{\rm cr}$, $\alpha$ approaches zero, then
the layer becomes stable against the DSI and unstable
for the gravitational instability only, as equation (55).
In the following paragraphs, we shall estimate $t_{\rm cr}$
and discuss evolution of the shocked slab.   

If the shock velocity decreases as $V_S \propto t^p ~(-1 < p < 0)$, 
we can estimate $t_{\rm cr}$ as follow. 
Column density at the time of $t$ is  estimated as 
\be
\sigma_0=q \rho_e R ~~~{\rm where}~~~R={V_S t \over p+1}.
\ee
Here, $q$ is the factor depending on the shape of the shock wave,
and, of course, $q=1$ is the case for the plane shock wave.
The distance of the 
shock surface from the origin is denoted as $R$. 
{}From the definition of $t_{\rm cr}$, $\alpha (t=t_{\rm cr})=1$,
and $P_S = \rho_e V_S^2$, we obtain the following relation for 
$t_{\rm cr}$;
\be
t_{\rm cr} = 
{(p+1) \beta ^{1/4} (1-\beta )^{1/4} \over 
q ( 2 \pi G \rho_e )^{1/2}}. 
\ee
Thus, $t_{\rm cr}$, 
which is the time for the onset of the gravitational instability,
does not depend on the initial shock velocity. 
Although $\beta $ is a function of time, 
the product of $\beta ^{1/4}$ and $(1-\beta )^{1/4}$
depends only very weakly on $\beta$. 
Indeed, for the range of $\beta = 0.1 - 0.9$, which is the enough range
for realistic situation,  $\beta ^{1/4} (1-\beta )^{1/4}$
has only a value in a range from 0.55 to 0.71. 
Thus, even for a realistic situation, 
the estimation of $t_{\rm cr}$ with the assumption of $\beta$ 
being constant is fairly valid.
That is to say, we estimate $t_{\rm cr}$ as being the order of 
the free-fall time of the external matter.

In our analysis, the effect of shell expansion is not included.
For the case of spherical expansion, the shell expansion effect 
may be important.  
The onset time of the gravitational instability
is estimated comparing the divergence rate of the 
expanding shells to the gravitational attraction rate 
for the expanding shells by Ostriker \& Cowie (1981) 
(in the cosmological case) 
and by McCray \& Kafatos (1987) (in the supershell case).  
Since we assume large compression ratio at the shock surface, 
the estimation for the snowplow phase (Eq. (22) of McCray \& Kafatos) 
should be compared with ours.  
The onset time for this case is somewhat shorter than 
the free-fall time of the external medium for the typical case. 
Thus, it may be correct that $t_{\rm cr}$ is somewhat shorter than 
the free-fall time of the external medium even for the expanding 
shell case, although more detailed investigation is necessary to
determine definitely.

Since the DSI saturates with transonic transverse flows 
in the shocked slab (Mac Low \& Norman 1993), 
only with the effect of the DSI the shocked slab hardly fragments. 
Thus, we should consider the coupling of the gravitational instability 
with the DSI for the fragmentation of the evolving shocked layer.
First, we consider the most unstable perturbation for the DSI
in an early stage ($t < t_{\rm cr}$).
Considering the constant wavelength, the perturbation grows by the DSI,
but it may be stable in the later stage as far as the perturbation
is linear (see figures 2-4).\footnote{Note that $\alpha$ decreases
and
$H$ increases (i.e. $K$ increases for a constant $k$) with time evolution.}
Next, we consider the perturbation being unstable for the gravitational
instability in the later stages.
That perturbation can grow by the DSI in the early stages.
Then, the DSI might be related to fragmentation process via
the gravity, which occurs in later stages.
By the way, importantly, in the intermediate stages,
the perturbation becomes stable against both instabilities
as shown in figure 3. Then, we find that
it is difficult for the DSI directly to connect with the gravitational
instability during the evolution of the slab.
The calculation by Yoshida \& Habe (1992) may suggest the stabilization
of the instabilities, that is,
the perturbation damps when the wavelength is in this stable range 
probably because of the evolution effect (e.g., Vishniac 1983,
Elmegreen 1989).
Thus, as far as the stable range is significant,
the effect of the DSI for the structure formation may not be
important. 
However, Yoshida \& Habe do not get the growing bending-mode for 
all parameters, although 
Mac Low \& Norman clearly present the evolution of the bending mode.
Thus, there is a possibility that numerical resolution of  Yoshida \&
Habe  is not enough. 
Further investigation including nonlinear stage
is necessary.

\section {Summary}

In this paper, we present the dispersion relation of the shocked
slab with self-gravity. According to it,
when the deceleration dominates the self-gravity,
the so-called DSI is efficient.
Their growth rates are the function of the scale length
which is determined from the efficiency of the deceleration.
Thus, we confirm previous researches qualitatively.
The slab becomes gravitationally unstable when the gravitational energy
is larger than the thermal energy of the layer which
is the result of the ram pressure at the shocked surface.
In other words, when the deceleration is inefficient,
the self-gravity works well.
Interestingly, even if the slab is gravitationally unstable 
in the long wave-length, it is not gravitationally unstable
in the short wave-length.
Instead of it, the DSI occurs. This suggests that
the smaller scale structure than the Jeans-length may be result 
of the DSI. 
The growth rate of the gravitational instability
depends on almost only $\alpha$, while  that of
the DSI does not depend only on $\alpha$ but also on $\beta$.
The growth time-scale of the both instabilities is
the order of the free-fall time for the slab density, while
the onset-time of the gravitational instability is
the order of the free-fall time of the external medium.
Thus, within our treatment, the fragmentation by the gravitational
instability becomes possible only after about the free-fall time of
the external medium.
The analytical new point in this paper is that we find 
stable modes in middle wave-number.
This is obtained since we are concerned with
the structure of the shocked slab with finite thickness (cf. Nishi 1992).
This stable range may suggest the ineffectiveness of  DSI for
the fragmentation of the shocked slab via the evolutional effects
(Vishniac 1983; Elmegreen 1989; Yoshida \& Habe 1992). 
Thus, we should determine the structure of the shocked layer when 
we must study the properties of the instabilities precisely.

\acknowledgements
We would like to thank the referee, Dr. M.-M. Mac Low, 
for his valuable comments.
We also thank Profs. H. Sato and  N. Sugiyama,
for their continuous encouragement.
HK appreciates
the computer maintenance by K. Yoshikawa.
This work is supported in part by Research Fellowships of the Japan
Society
for the Promotion of Science for Young Scientists, No.2780 and 03926
(HK),
and by the Japanese Grant-in-Aid
for Scientific Research on Priority Areas (No. 10147105) (RN) and
Grant-in-Aid for Scientific Research of the Ministry of Education,
Science, Sports, and Culture of Japan, No. 08740170 (RN).

\newpage
 
\centerline{\bf FIGURE CAPTIONS}

Fig.1 ---
The numerical expression of (\ref{threeeight}).
The growth rate is normalized by $c/H$,
that is, $\Omega = \omega / (c/H)$.
In the figure, $\Omega_r $ is a real part of $\Omega$,
while $\Omega_i $ is an imaginary part of $\Omega$. 
The wave-number is expressed as $K = kH$.
The positive real parts are shown in (a), while
the dotted line corresponds to the plus sign of (\ref{threeeight}) 
and the solid line to its minus sign. 
In (b), the two imaginary parts of both the modes coincide.

Fig.2 ---
The numerical expression of (\ref{gthreeseven}).
The growth rate is normalized by $(2\pi G \rho _s)^{0.5}$.
We draw $\Omega_r$ as a real part of $\Omega$,
while $\Omega_i $ as an imaginary part of $\Omega$.
The wave-number is expressed as $K = kH$.
We adopt $\alpha = 1.2$  and 
$\beta=$ 0.2(solid), 0.4(dash), 0.6(dot) and 0.8(dot-dash).
Only DSI modes are found for the four examples.

Fig.3 ---
The same figure of Fig. 2 except the value of $\alpha = 0.7$.
We also adopt $\beta=$ 0.2(solid), 0.4(dash), 0.6(dot) 
and 0.8(dot-dash).
In small $K$ we find the gravitational instability, while DSI occurs
in large $K$. In the middle of $K$, we find just the stable modes.
The case of $\beta =0.8$ does not have DSI even in large range of $K$.
Of course, $\Omega_r =0$ for the gravitationally unstable modes.

Fig.4 ---
The same figure of Fig. 2 except the value of $\tilde \alpha,~\alpha=0.0$.
We find the dispersion relation given in Goldreich and Lynden-Bell (1965).
The solid line shows the gravitational instability, while
the dotted line is only for the gravitationally modified sound mode.

Fig.5 ---
The one set of the evolution of the linear displacements are displayed.
Here,  $\eta (\equiv z_{s 1} - 0)$ and $\zeta (\equiv z_{t 1} - H)$, then
we show $z_{s 1}$ as dotted lines and $z_{t 1}$ as solid lines.
Each of the displayed time is (a)$0.0$, 
(b)$1.0 \cdot \delta t$, and (c)$2.0 \cdot \delta t$.
The unit of the length is $H$.
The bending property of the shocked layer for DSI is found.

Fig.6 ---
The same figure of figure 5.
The displayed times are 
(a)$0.0$, (b)$1.0 \cdot \delta t$, and (c)$2.0 \cdot \delta t$.
The gravitational-unstable slab is presented.

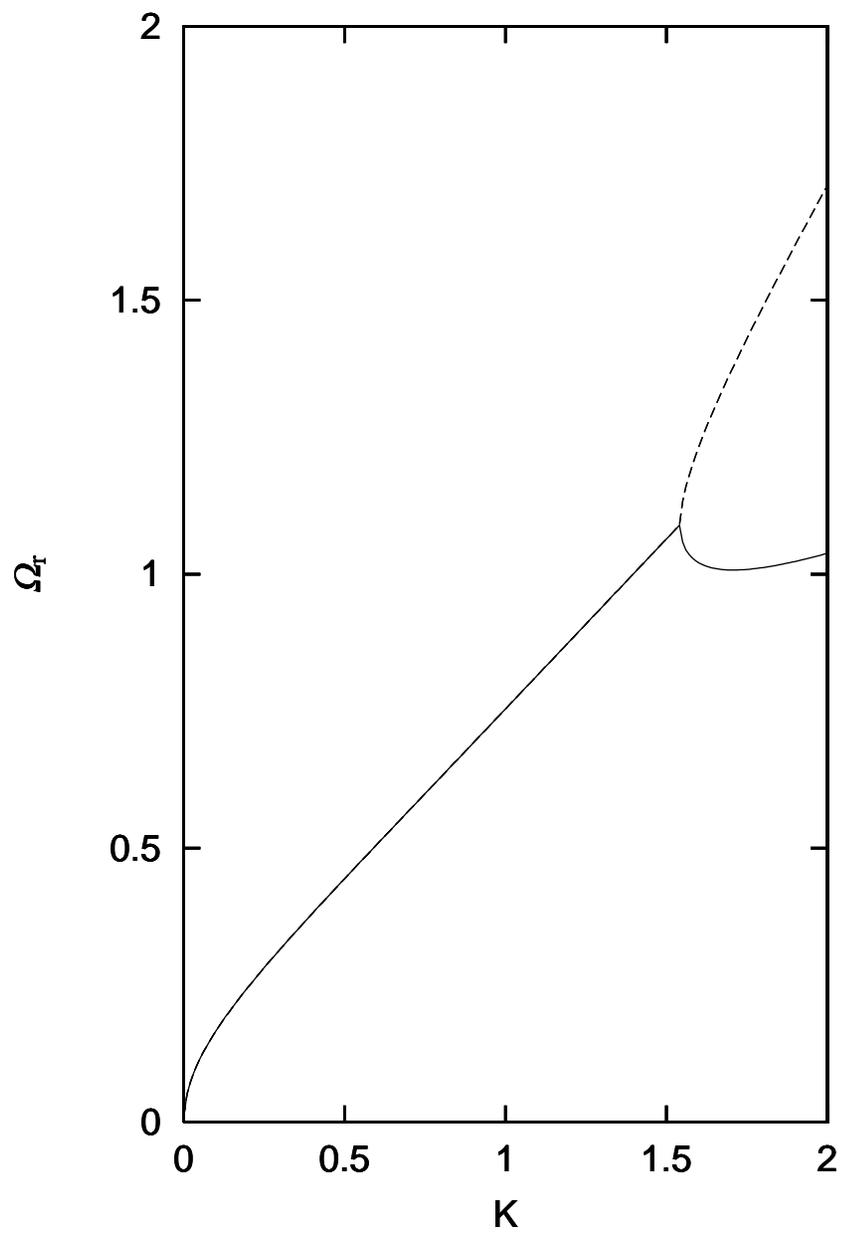
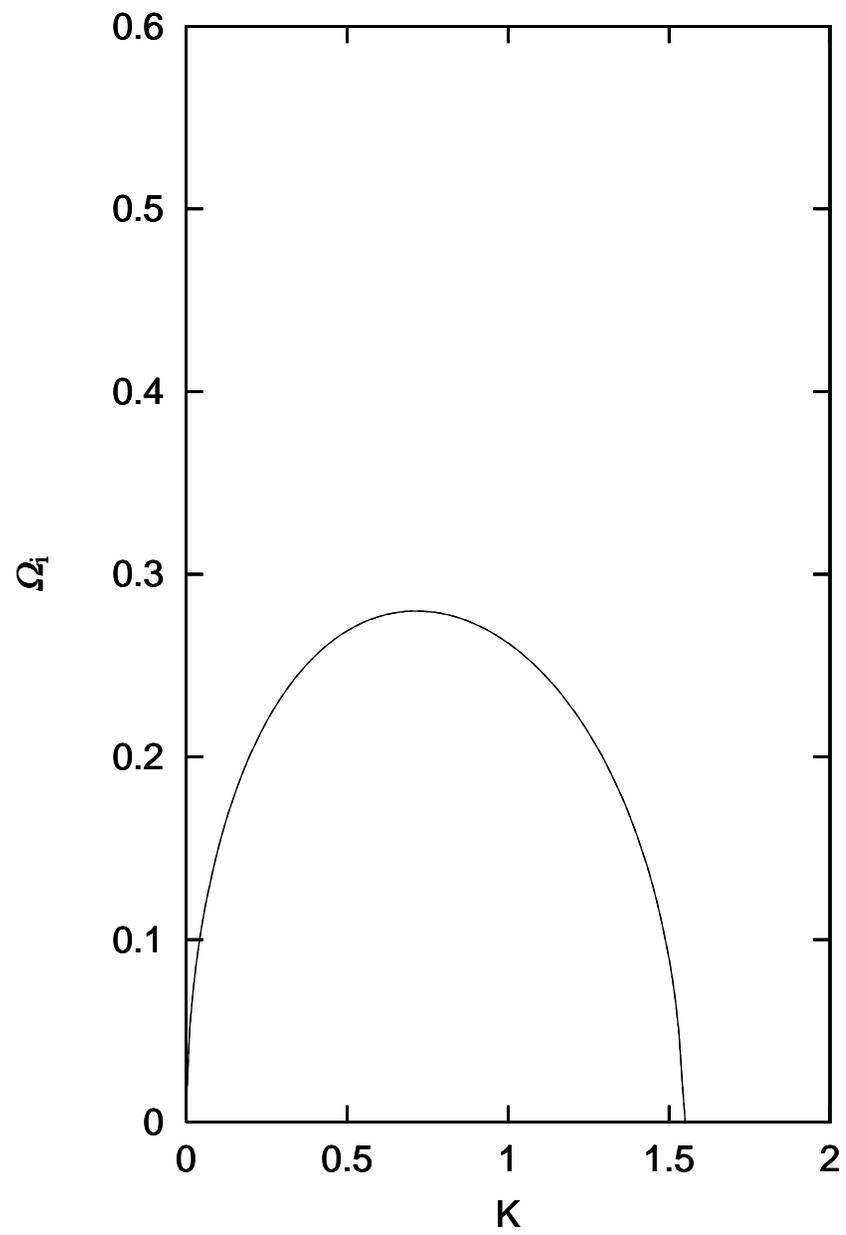
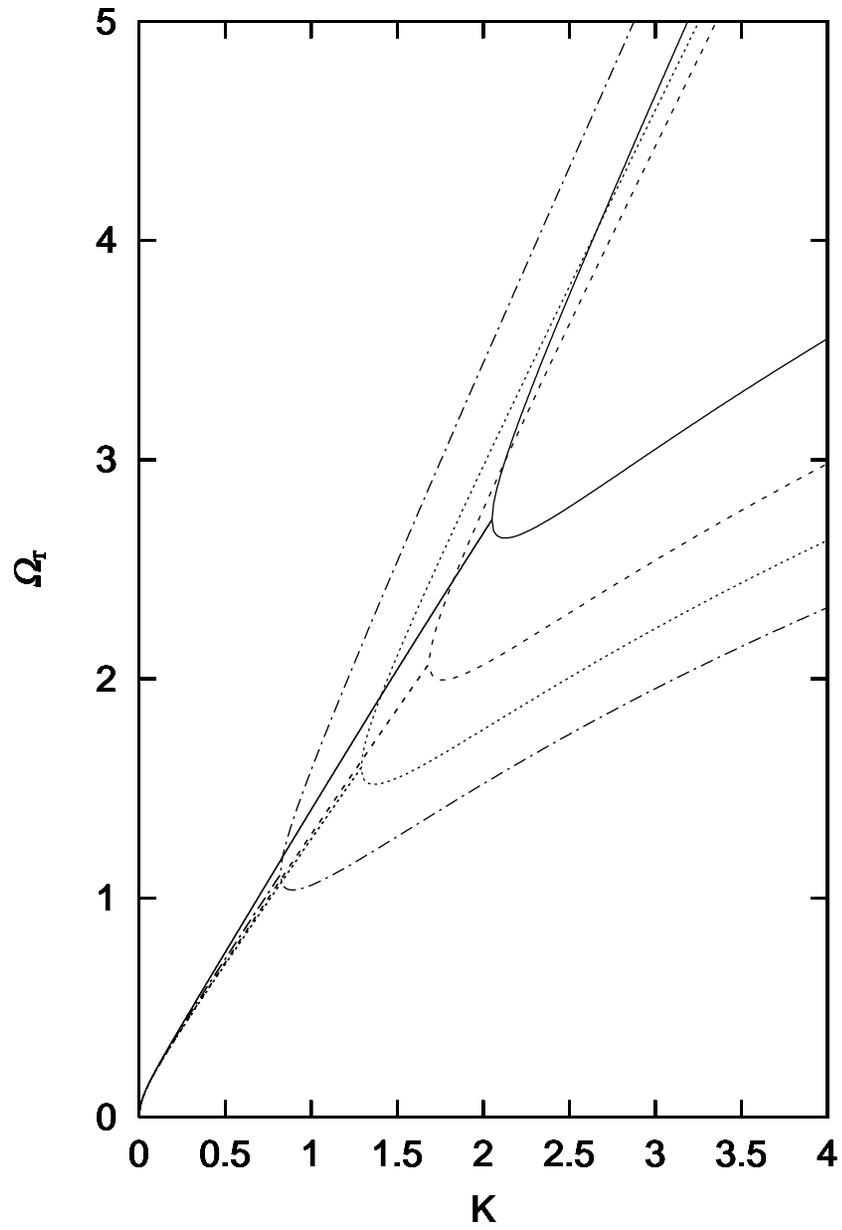
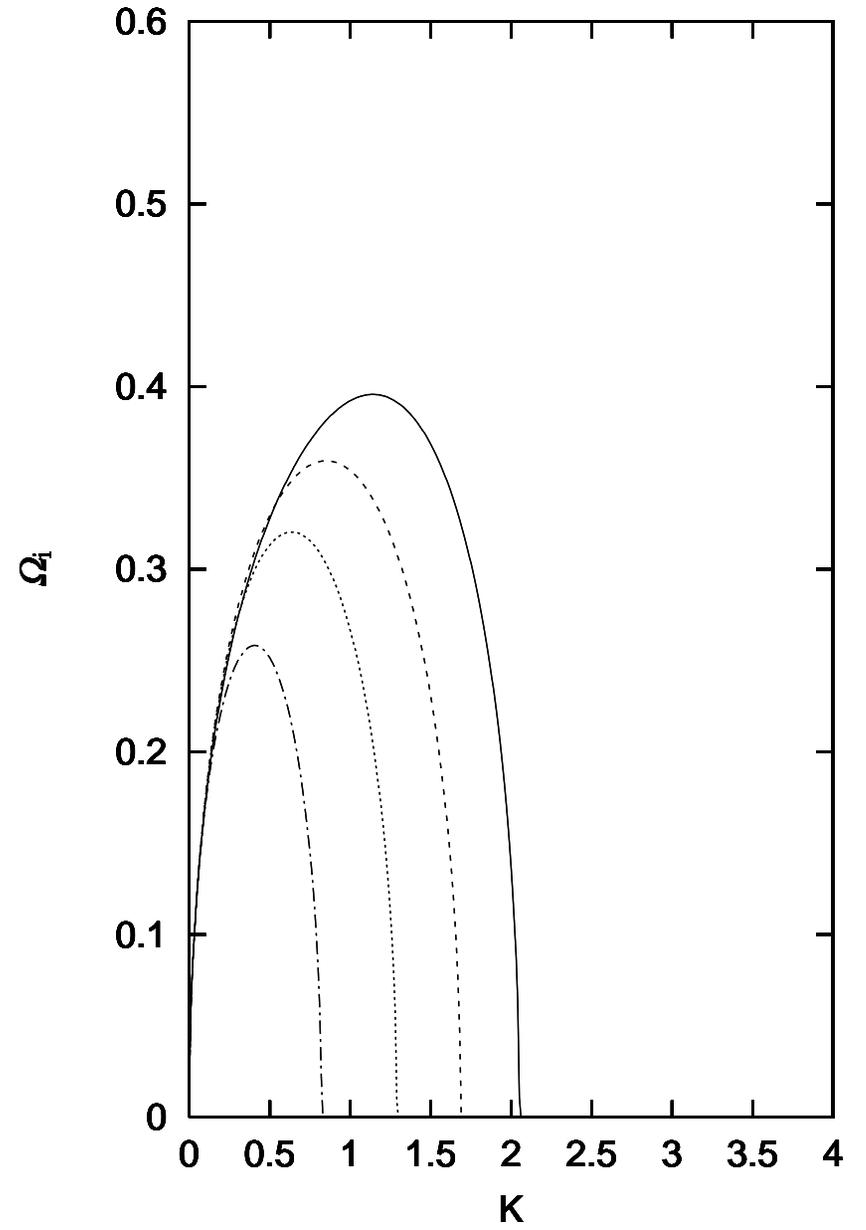
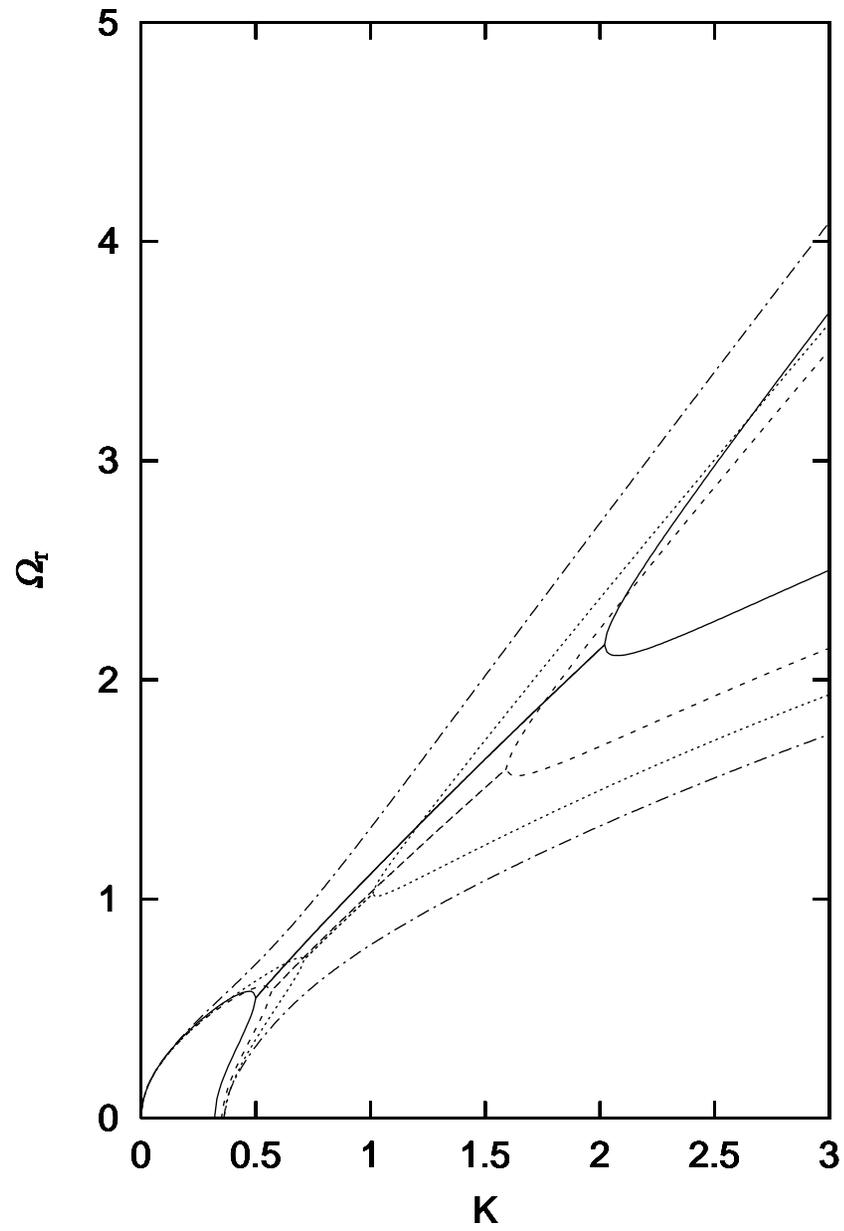
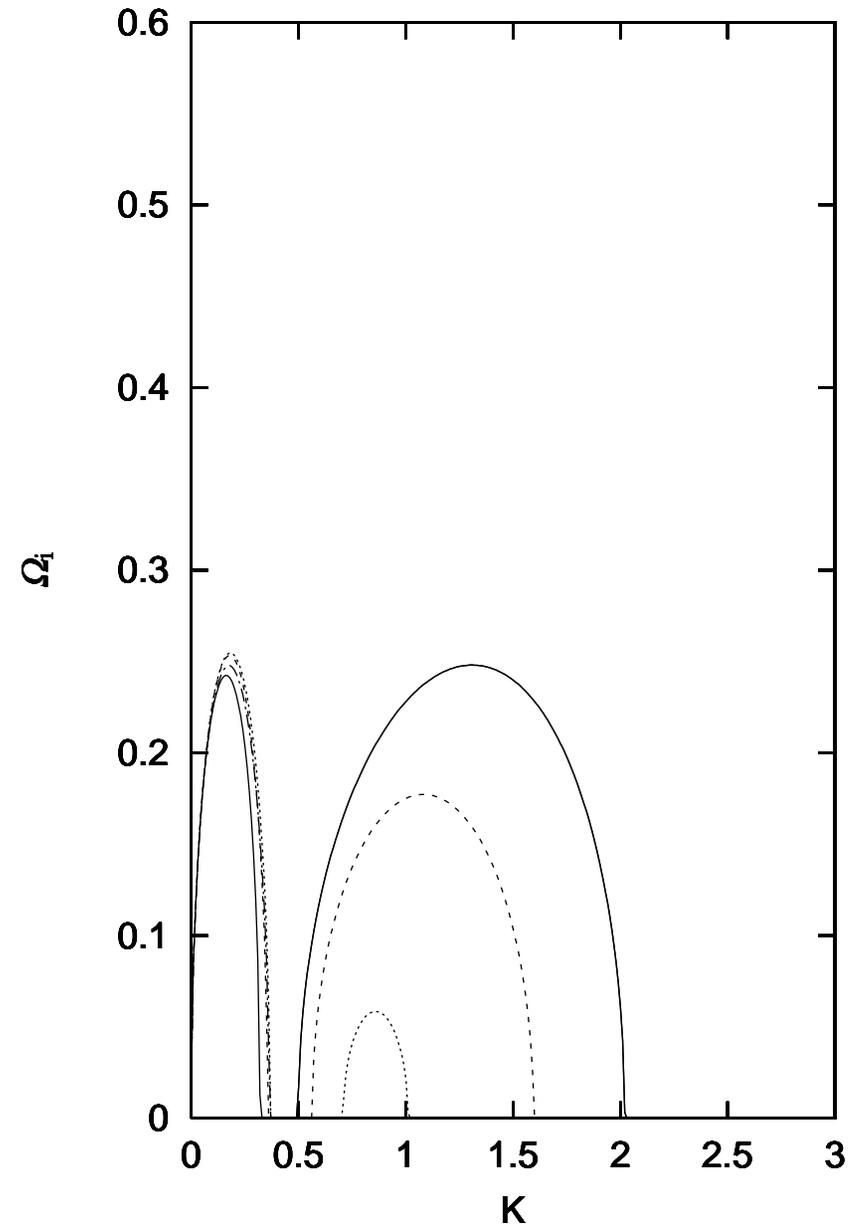
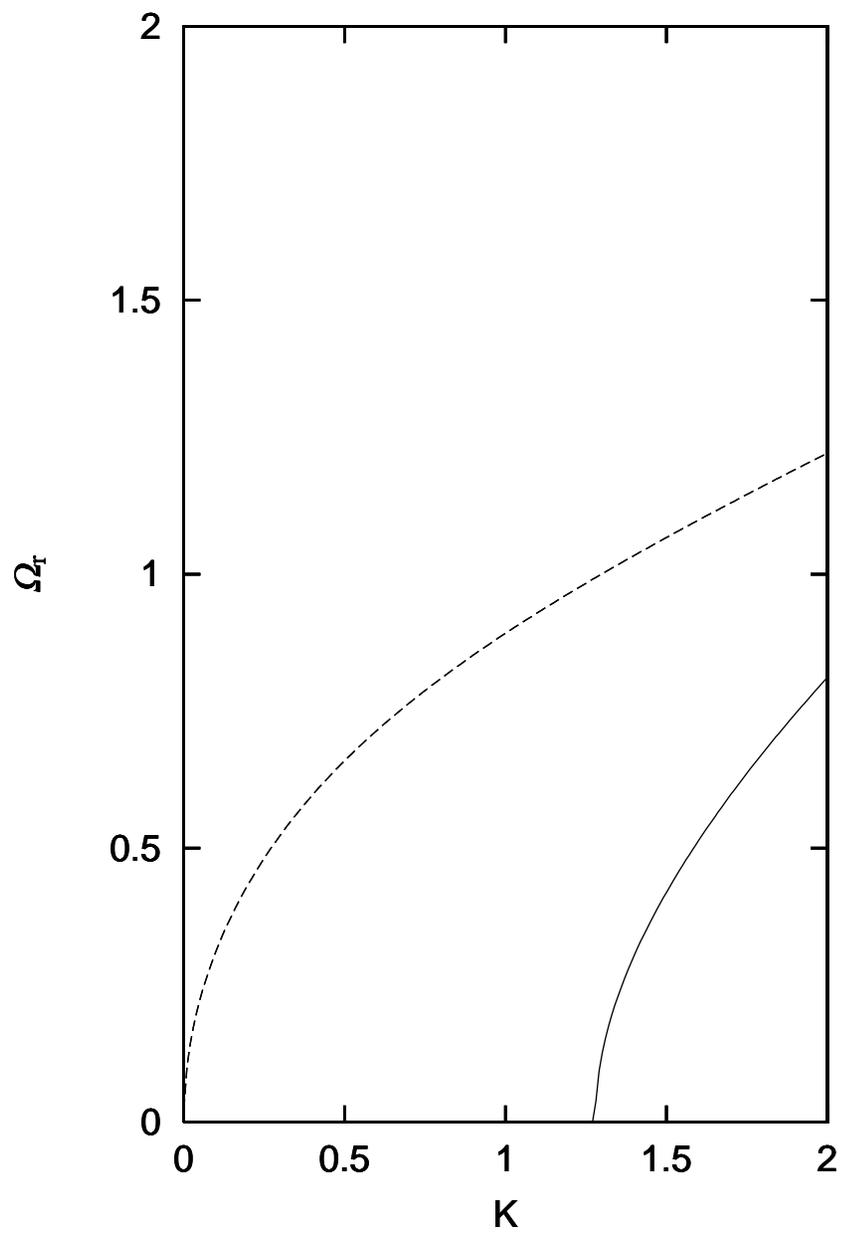
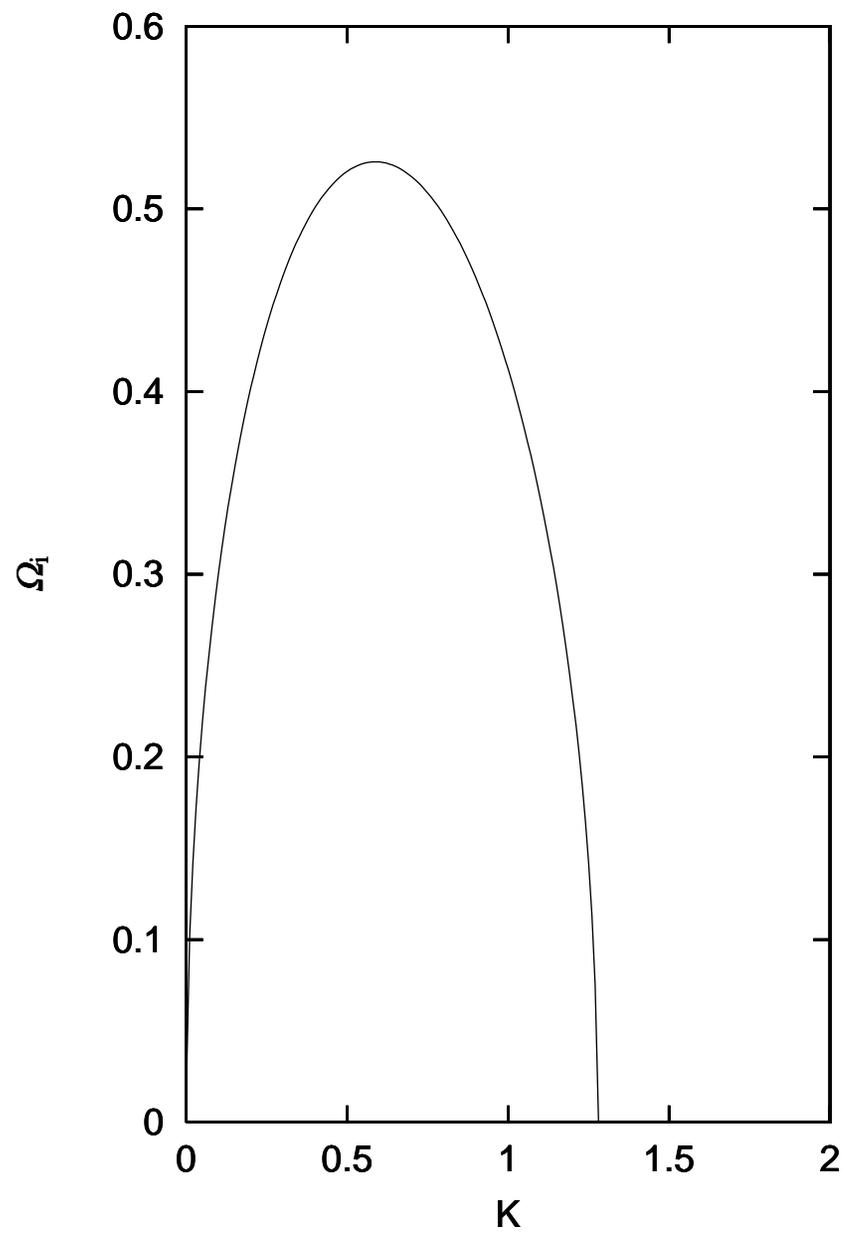
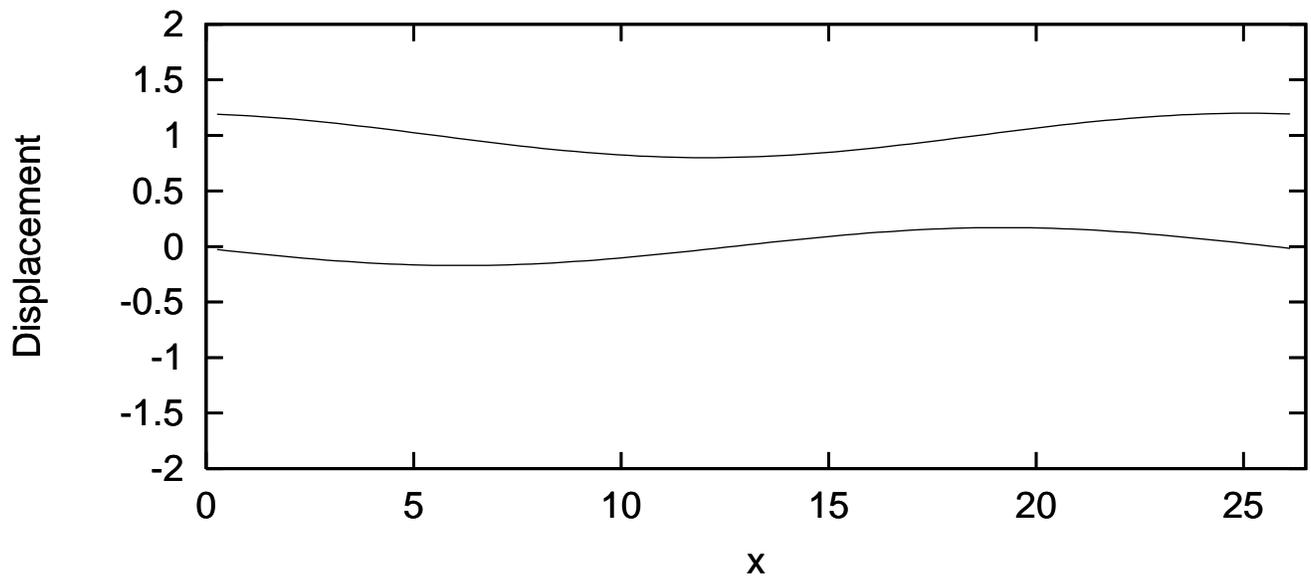
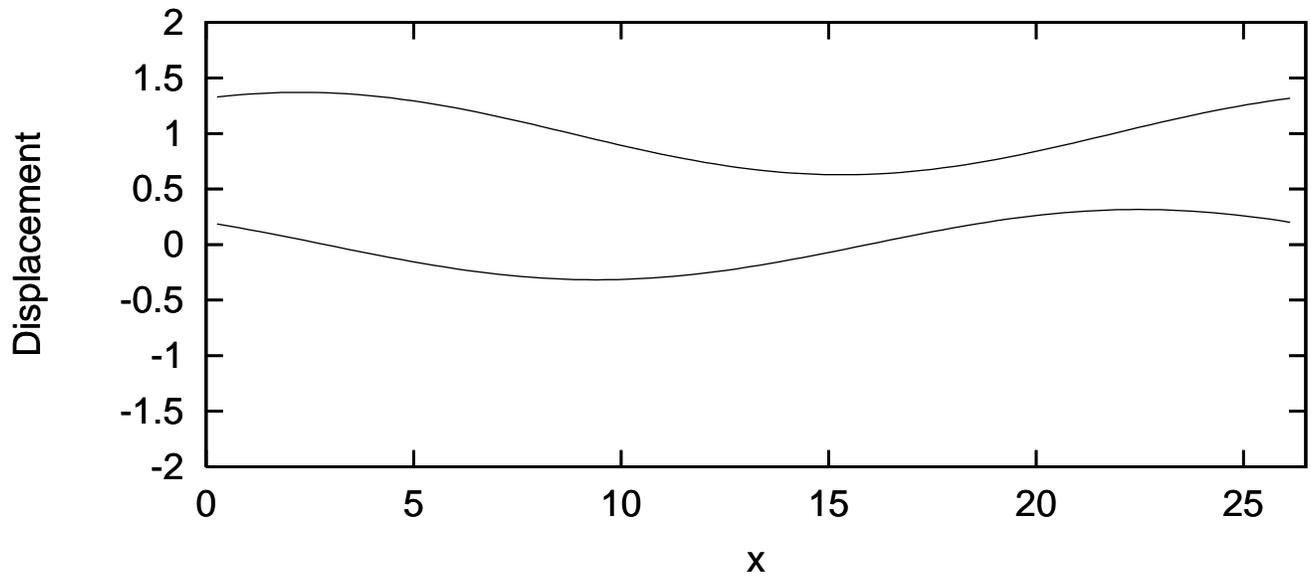
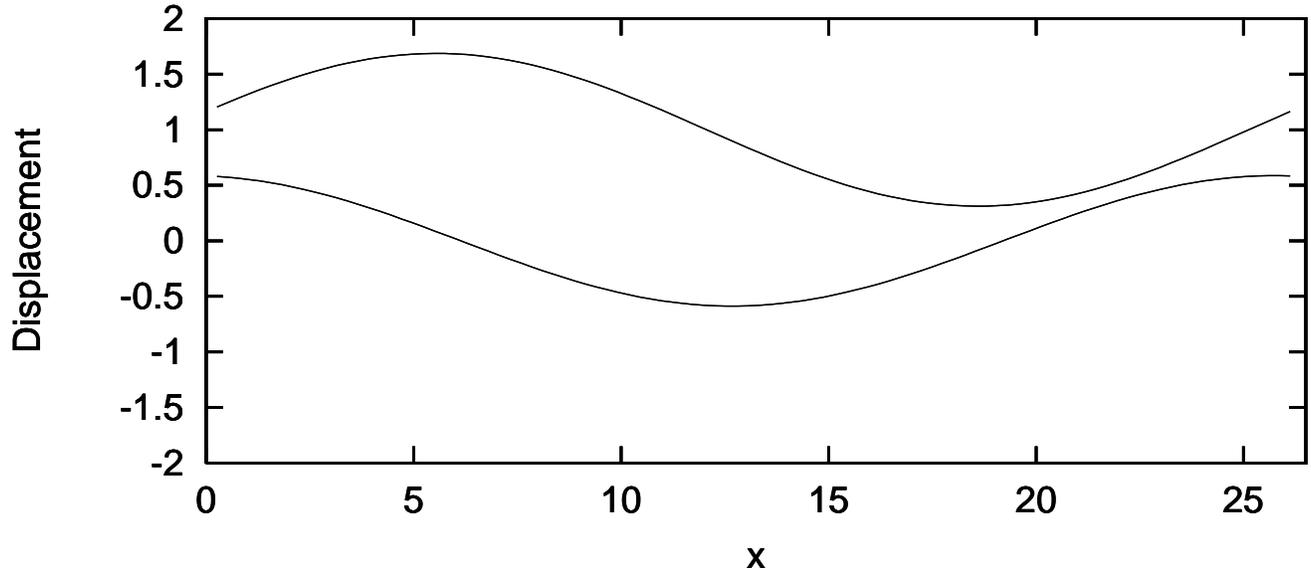
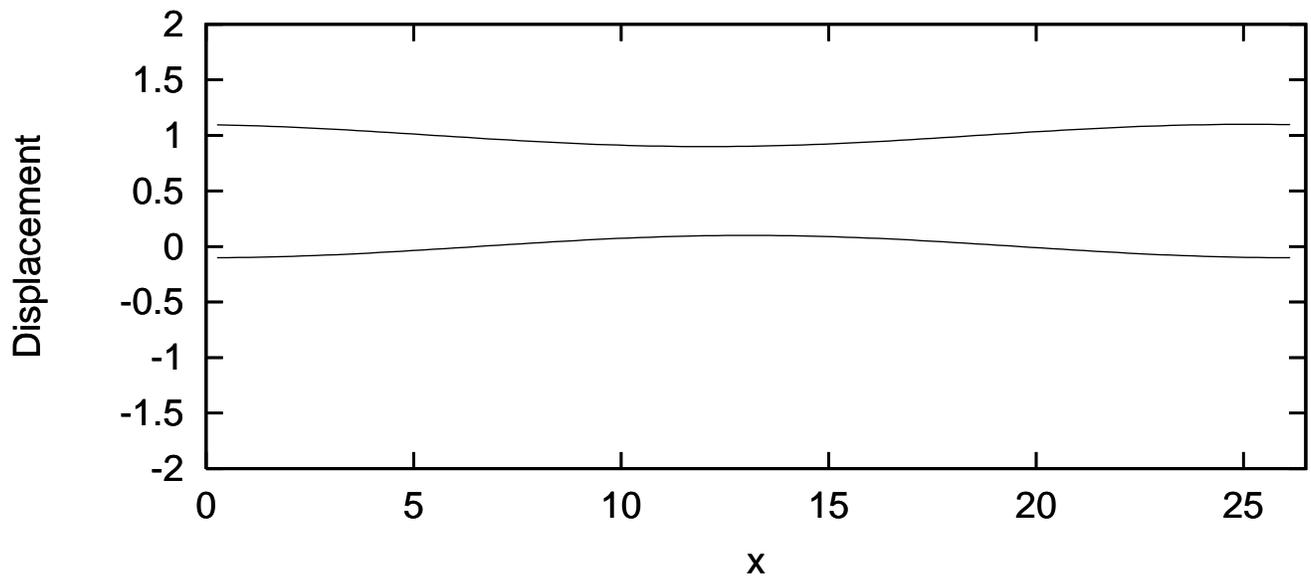
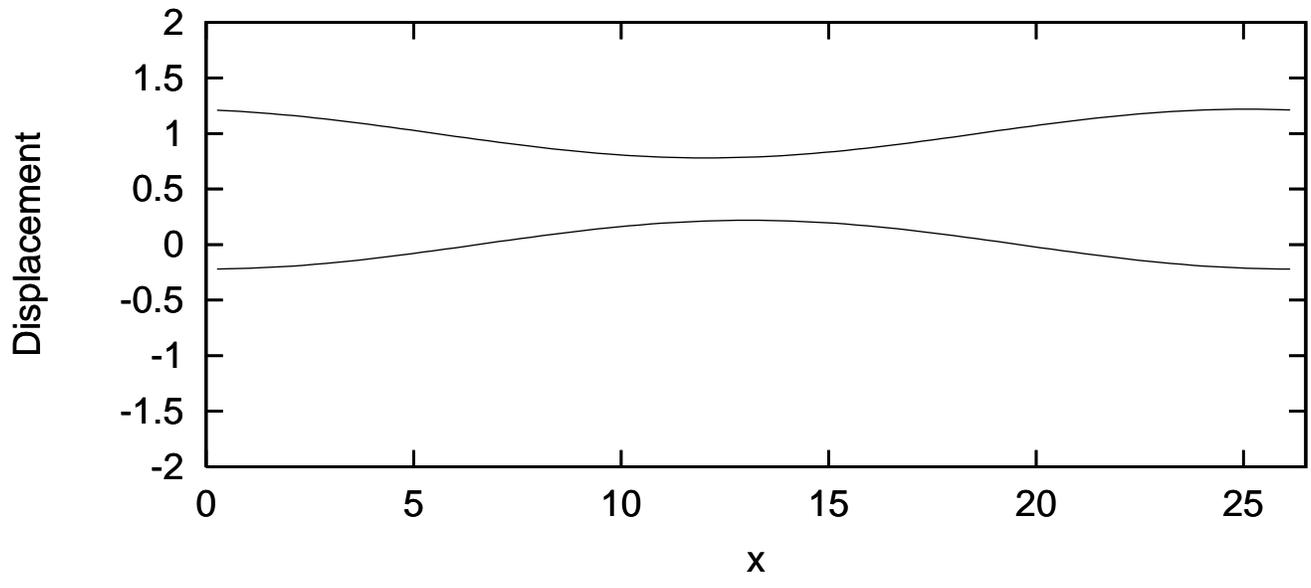
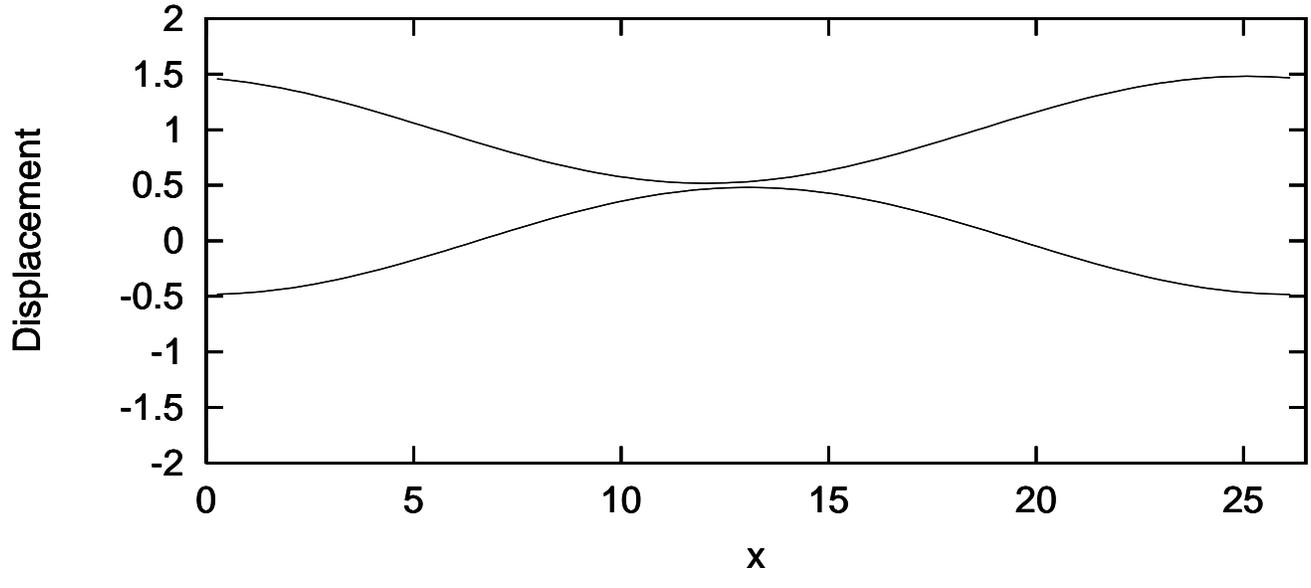
\end{document}